%% file: paper.tex
\documentclass[preprint2, times, twocolappendix]{aastex702}

\input{tex/build/macros/project.tex}

\input{tex/src/preamble-maneage.tex}

\input{tex/src/preamble-project.tex}

\begin{document}

\title{\projecttitle}

\author[orcid=0000-0003-1710-6613,gname=Mohammad,sname=Akhlaghi]{Mohammad Akhlaghi}
\affiliation{Centro de Estudios de Física del Cosmos de Aragón (CEFCA), Plaza San Juan, 1, E-44001, Teruel, Spain}
\email{mohammad@akhlaghi.org}

\author[orcid=0000-0002-1071-9508,gname=Héctor,sname='Vives Arias']{Héctor Vives Arias}
\affiliation{Centro de Estudios de Física del Cosmos de Aragón (CEFCA), Plaza San Juan, 1, E-44001, Teruel, Spain}
\email{hvives@cefca.es}

\author[orcid=0000-0002-5953-4491,gname=Pablo,sname=Renard]{Pablo Renard}
\affiliation{Institute of Space Sciences (ICE, CSIC), Campus UAB, Carrer de Can Magrans, s/n, 08193 Barcelona, Spain}
\affiliation{Institut d\'~Estudis Espacials de Catalunya (IEEC), 08034 Barcelona, Spain}
\email{renard@ieec.cat}

\author[orcid=0000-0003-3135-2191,gname=Héctor,sname='Vázquez Ramió']{Héctor Vázquez Ramió}
\affiliation{Centro de Estudios de Física del Cosmos de Aragón (CEFCA), Plaza San Juan, 1, E-44001, Teruel, Spain}
\email{hvr@cefca.es}

\author[orcid=0000-0002-6220-7133,gname=Raúl,sname=Infante-Sainz]{Raúl Infante-Sainz}
\affiliation{Centro de Estudios de Física del Cosmos de Aragón (CEFCA), Plaza San Juan, 1, E-44001, Teruel, Spain}
\email{rinfante@cefca.es}

\begin{abstract}
  \noindent
  Deep astronomical images are produced by the coadding (also called ``stacking'') of several individual exposures.
  However, individual exposures are often plagued with many artifacts which will reduce the quality of the deep image if not properly removed.
  Most are removed during reduction/calibration, but the more diffuse may stay.
  To address this problem, the following new operators were introduced to the Arithmetic program in Gnuastro 0.23: \texttt{madclip-maskfilled}, \texttt{sigclip-maskfilled} and \texttt{collapse-*clip-fill-*}.
  They are able to mask many such outliers successfully by using all the exposures to find outlying pixels in each exposure and exploiting their contiguity.
  This research note is reproducible with Maneage, on the Git commit \projectversion.
\end{abstract}


\section{Introduction}\label{sec:intro}

When building the final deep image from individual exposures, it is critical that artifacts are excluded \citep[see][and references therein]{desai16}.
The more diffuse and extended the artifact are on each exposure, the harder their removal is in the reduction pipeline and its calibration steps.
The last line of defense is in the pipeline's coadding step: where we have multiple images covering the same location on the sky.

\section{Existing solutions}\label{sec:analysis}

\begin{figure}[t]
  \vspace{1cm}
  \ifdefined\makepdf%
    \tikzsetnextfilename{figure}%
    \input{tex/src/figure.tex}%
  \else
    \includegraphics[width=\linewidth]{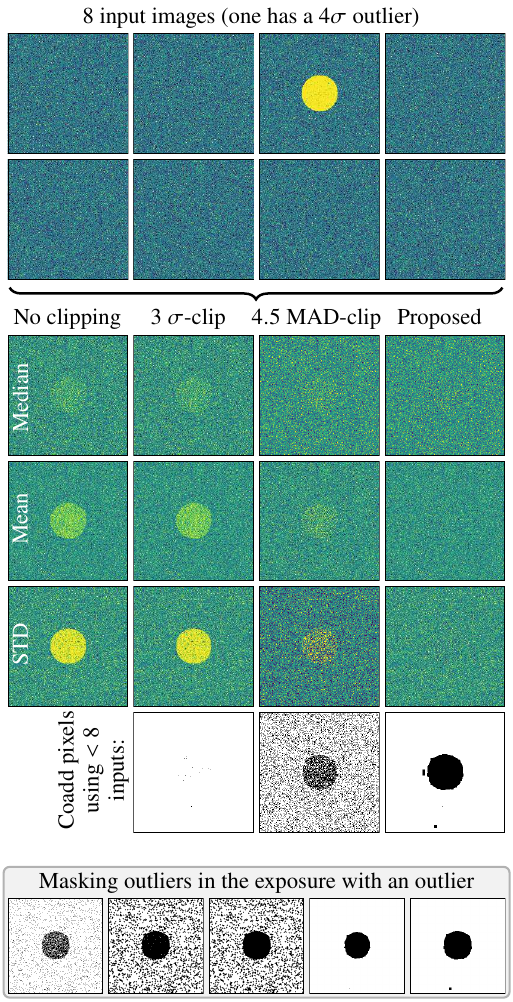}
  \fi

  \vspace{-7mm}
  \caption{\label{figure} Outlier rejection.
    The top two rows show the $\mocknumber$ individual exposures that are used as input.
    The next four rows show coadds with various statistical measures.
    The columns of the coadd rows show the result of coadding with different operators (the columns are respectively: no clipping, 3 $\sigma$-clipping, 4.5 MAD clipping and the method mentioned here).
    The bottom row (in a box) shows the steps of the newly introduced method only on a single exposure (the one with an outlier).}
\end{figure}

One solution is to use ``robust'' coadding operators like the median, sigma-clipping, or median absolute deviation (MAD) clipping.
However, they are not as robust when dealing with extended signal in a noisy image.
For example Figure \ref{figure} shows $\mocknumber$ mock images with pure noise.
But one has an outlier in it: a flat circle where each pixel has a significance of $\mockpixsn\sigma$ in relation to the background noise.

To help in visual comparison, the Median, Mean and STD coadd scales of Figure \ref{figure} are set such that the noise is the same color in all of them.
From the first column (``No clipping''), we see how the standard deviation (STD) coadd (where every pixel is the standard deviation of the same pixel in all eight inputs) is the most strongly affected by that outlier: the circle is brightest.
The effect of the outlier is smaller in the mean and even smaller in the median, but it does not disappear.

To better understand why this happens, here are the 8 values in the central pixel of this image (the noise in all the images has a standard deviation $\mocksigma$ units): $\mockpixdemoimgone$, $\mockpixdemoimgtwo$, $\mockpixdemoimgthree$, $\mockpixdemoimgfour$, $\mockpixdemoimgfive$, $\mockpixdemoimgsix$, $\mockpixdemoimgseven$ and $\mockpixdemoimgeight$.
The standard deviation of these values without and with the outlier dataset are respectively: $\mockpixdemostdgood$ (similar to the ground-truth of $\mocksigma$) and $\mockpixdemostdraw$ (almost 50\% higher).

The next two columns show the coadds after clipping (in a Gaussian distribution, $4.5 \rm{MAD}\approx3\sigma$).
In particular, $\sigma$-clipping \citep[the most common way to reject outliers, see Section 2.10.2 of][]{gnuastro24} has barely had any effect.
The reason can be displayed with the values of the same pixel of the previous paragraph: the median ($m$) of those 8 values is $\mockpixdemomedian$, with $\sigma=\mockpixdemostd$; therefore the clipping threshold ($m+\sigclipsigma\sigma$) is $\mockpixdemothresh$ and the outlier above ($\mockpixdemoimgthree$) is not clipped.
Very few pixels will be clipped when $\sigma$ itself is so strongly affected.

MAD-clipping \citep[Section 2.10.3 of][]{gnuastro24} has removed the object more successfully because the MAD is more robust to outliers than the standard deviation.
However, we can see that the purity of MAD clipping is very poor: many non-outlier pixels are discarded, with a visible effect on the noise/depth.

\section{Solution}

The proposed solution's steps are applied to each exposure and are listed below.
The box at the bottom of Figure \ref{figure} shows each listed step for the exposure with an outlier.

\begin{enumerate}
\item Clipping is done once on all the inputs and all clipped pixels are given a value of one, while the rest are set to zero; producing a binary image.
  MAD-clipping is used here and we see that many of the outlier pixels get a value of one, but due to noise, it is very porous.
\item The binary image of the previous step is dilated with 4-connected neighbors \citep[see Section 3.1.4 of][]{gnuastro}.
  This causes many of the randomly missed pixels of the circle to get a value of one, but also makes the false pixels (that were not an outlier) a little thicker.
\item The holes (any group of zero-valued pixel that are fully surrounded by one-valued pixels in all 8 neighbors) are filled (get a value of one).
  This ensures that all mixed pixels enclosed within the outlying region are also masked later.
\item Two 8-connected erosions are applied to remove all the false positives. This image is the first with a clean background: all the false positives (non-outliers) are removed in this step because they were randomly positioned.
\item Two 8-connected dilations are applied so the edges of the true outliers that were eroded in the previous step are covered again.
  All the pixels with a value of one in this exposure are set to NaN (not-a-number).
\end{enumerate}

Once all the steps above are done on every exposure they are coadded with any coadding operator (like the mean), and the result is shown in the last (``Proposed'') column of Figure \ref{figure}.
We see that there is no footprint of the outlier in that column.
As in any method, this solution also has its limits: when the diffuse signal is not bright enough for the dilation and filling of holes to catch its contiguity, the outlier will be missed.
However, for outliers/artifacts brighter than this limit it will be able to remove them in the coadd.

The steps above are implemented in GNU Astronomy Utilities (Gnuastro) since version 0.23 as the following two new operators of its Arithmetic program: \texttt{madclip-maskfilled} and \texttt{sigclip-maskfilled}.
This concept has also been implemented in 1D as \texttt{collapse-*clip-fill-*}; for example when collapsing pixel rows as in Figure 4 of \citet{nasim}.
A complete tutorial has also been written for the optimal usage of these operators in Section 2.10 (called \emph{Clipping outliers}) of \citet{gnuastro24}.

Gnuastro is an actively developed software; if you are using later versions of Gnuastro be sure to check the \texttt{NEWS} file for any change after this publication.

\section{Acknowledgments}

This project was developed in the reproducible framework of Maneage \citep[\emph{Man}aging data lin\emph{eage},][latest Maneage commit \maneageversion{}, from \maneagedate]{maneage}.
It was built on an {\machinearchitecture} machine with {\machinebyteorder} byte-order and created from commit {\projectversion} of its source (branch \texttt{\small{\projectgitbranch}} of \url{\projectgitrepo}).
The source code is archived on Software Heritage\footnote{\href{https://archive.softwareheritage.org/swh:1:dir:84092a4318c15846b51fe6124dac5ba47fd191b2;origin=https://codeberg.org/gnuastro/papers;visit=swh:1:snp:dafbd00819bfa33f5cc773783ad044b5b7f3177f;anchor=swh:1:rev:f067b60989f6940a6e163c0345b478fe0917f71a}{\texttt{swh:1:dir:84092a4318c15846b51fe6124dac5ba47fd191b2}}} and \href{\projectdoizenodo}{Zenodo.\projectzenodoid} for longevity.

We are grateful to David Muniesa, Alberto Moreno Signes and Stergios Amarantidis as well as grants PID2021-124918NA-C43 and PID2024-162229NB-I00 funded by MICIU/AEI/10.13039/501100011033 and by “ERDF A way of making Europe” (ERDF/EU), by the European Union.

\bibliography{references}{}
\bibliographystyle{aasjournalv7.1}

\appendix

\section{Software acknowledgment}
\label{appendix:software}
\input{tex/build/macros/dependencies.tex}

\end{document}

%% file: tex/build/macros/project.tex
\input{tex/build/macros/hardware-parameters.tex}

\input{tex/build/macros/initialize.tex}
\input{tex/build/macros/mock-inputs.tex}
\input{tex/build/macros/coadd.tex}

%% file: tex/build/macros/hardware-parameters.tex
\newcommand{\machinearchitecture}{x86\_64}
\newcommand{\machinebyteorder}{Little Endian}

%% file: tex/build/macros/initialize.tex
\newcommand{\projecttitle}{Gnuastro: removing extended/diffuse outliers while coadding}
\newcommand{\projectkeywords}{Astronomy data reduction, Astronomy image processing, Astronomy software, Coadding, Outliers}
\newcommand{\projectzenodoid}{22676761}
\newcommand{\projectversion}{bf3815a}
\newcommand{\projectdoizenodo}{https://doi.org/10.5281/zenodo.22676761}
\newcommand{\projectgitrepo}{https://codeberg.org/gnuastro/papers}
\newcommand{\projectgitbranch}{maskfilled}
\newcommand{\projectcopyrightowner}{Mohammad Akhlaghi <mohammad@akhlaghi.org>}

\newcommand{\maneagedate}{29 Jul 2026}
\newcommand{\maneageversion}{6f0ecbe}

%% file: tex/build/macros/mock-inputs.tex
\newcommand{\mocksigma}{10}
\newcommand{\mocknumber}{8}
\newcommand{\mockpixsn}{4}
\newcommand{\mockpixdemoimgone}{5.35}
\newcommand{\mockpixdemoimgtwo}{-3.38}
\newcommand{\mockpixdemoimgthree}{41.54}
\newcommand{\mockpixdemoimgfour}{22.15}
\newcommand{\mockpixdemoimgfive}{10.27}
\newcommand{\mockpixdemoimgsix}{0.65}
\newcommand{\mockpixdemoimgseven}{19.01}
\newcommand{\mockpixdemoimgeight}{-2.25}
\newcommand{\mockpixdemomedian}{7.81}
\newcommand{\mockpixdemostd}{14.32}
\newcommand{\mockpixdemothresh}{50.77}

%% file: tex/build/macros/coadd.tex
\newcommand{\madclipmad}{4.5}

\newcommand{\sigclipsigma}{3}
\newcommand{\mockpixdemostdraw}{14.32}
\newcommand{\mockpixdemostdgood}{9.41}

%% file: tex/src/preamble-maneage.tex
\ifdefined\highlightnew

\else

\fi

\ifdefined\highlightnotes
\newcommand{\tonote}[1]{\textcolor{red!60!black}{[#1]}}
\else
\newcommand{\tonote}[1]{{}}
\fi

%% file: tex/src/preamble-project.tex
\usepackage{graphicx}

\usepackage{xcolor}
\color{black} 
\definecolor{DarkBlue}{RGB}{0,0,90}

\hypersetup{
    pdftitle={\projecttitle},
    pdfauthor={\projectcopyrightowner},
    pdfsubject={\projectgitrepo{} (commit \projectversion)},
    pdfkeywords={\projectkeywords}
}

\input{tex/src/preamble-pgfplots.tex}

\shorttitle{\projecttitle}
\shortauthors{Akhlaghi, M}

\newcommand{\imginfig}[4]{
  \node[anchor=south west] (img) at (#2\linewidth,#3\linewidth)
       {\includegraphics[width=#4\linewidth]
                        {tex/build/figures/#1.pdf}}}

\newcommand{\rotatedtxt}[4]{
  \node[anchor=north, rotate=90] at (#2\linewidth,#3\linewidth){\textcolor{#4}{#1}};
}

%% file: tex/src/preamble-pgfplots.tex
\usepackage{tikz}
\usetikzlibrary{external}
\tikzsetexternalprefix{tex/tikz/}

\usepackage{pgfplots}
\pgfplotsset{compat=newest}
\usepgfplotslibrary{groupplots}
\pgfplotsset{
  axis line style={thick},
  tick style={semithick},
  tick label style = {font=\footnotesize},
  every axis label = {font=\footnotesize},
  legend style = {font=\footnotesize},
  label style = {font=\footnotesize}
  }

%% file: tex/src/figure.tex
\begin{tikzpicture}

  \node[anchor=south] at (0.5\linewidth,0.25\linewidth)
       {8 input images (one has a $\mockpixsn\sigma$ outlier)};
  \imginfig{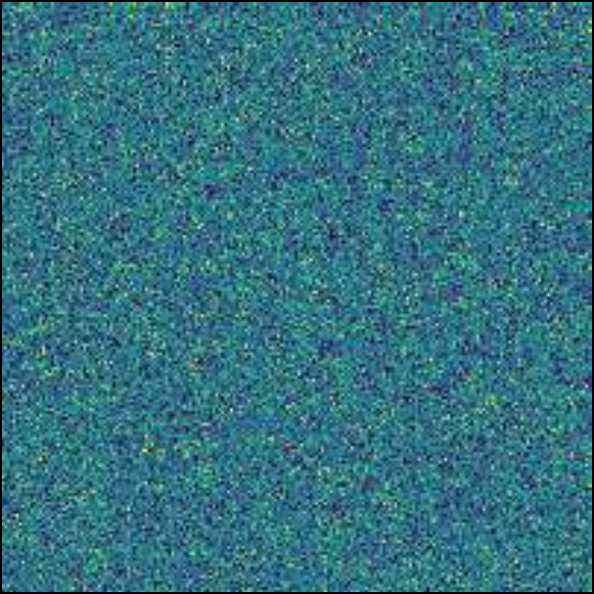}{0.00}{+0.00}{0.24};
  \imginfig{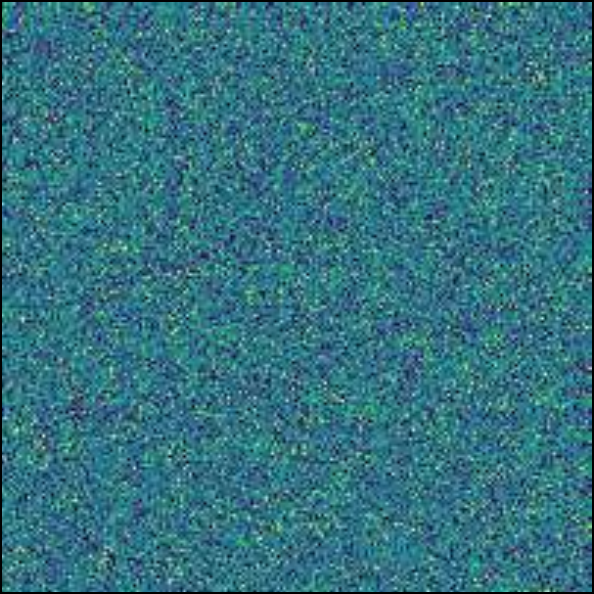}{0.25}{+0.00}{0.24};
  \imginfig{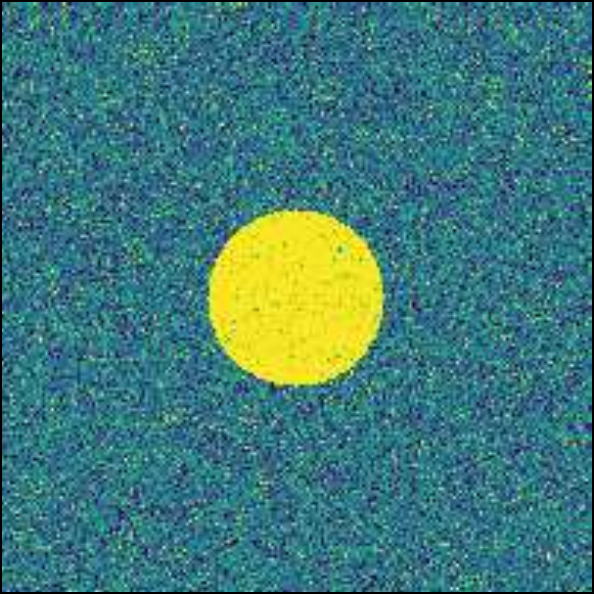}{0.50}{+0.00}{0.24};
  \imginfig{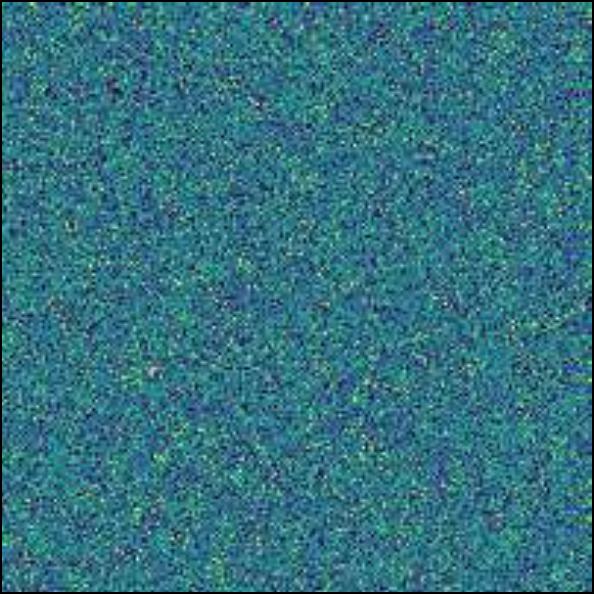}{0.75}{+0.00}{0.24};
  \imginfig{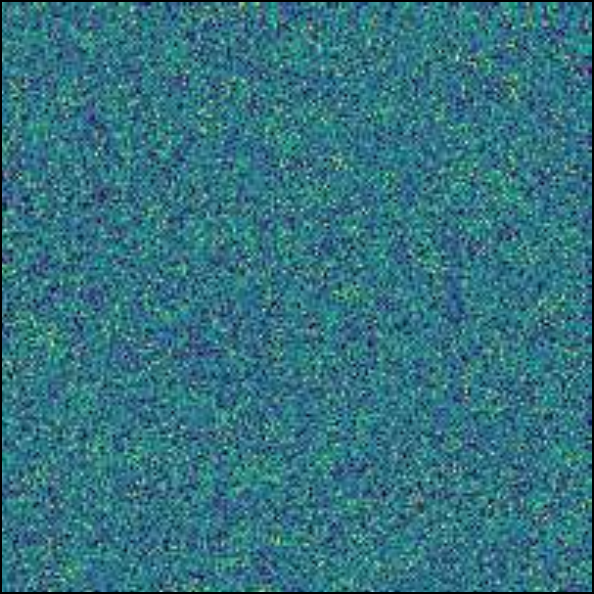}{0.00}{-0.25}{0.24};
  \imginfig{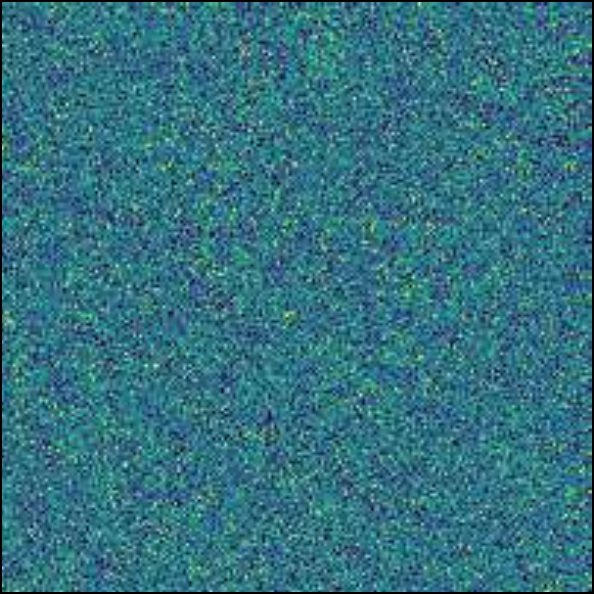}{0.25}{-0.25}{0.24};
  \imginfig{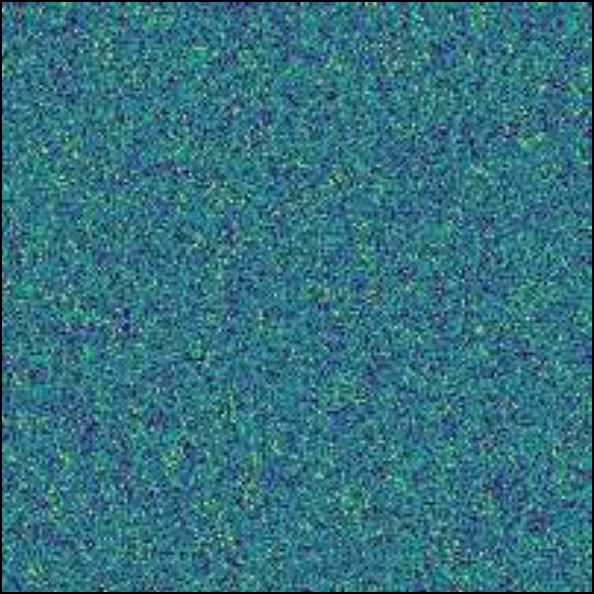}{0.50}{-0.25}{0.24};
  \imginfig{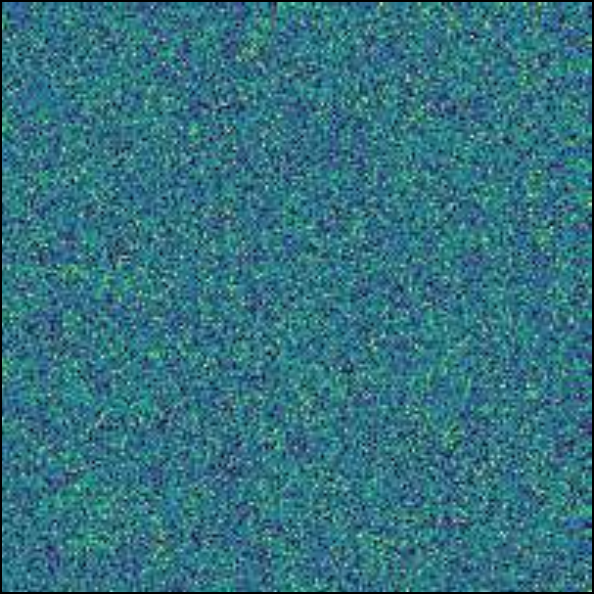}{0.75}{-0.25}{0.24};

  \draw [very thick, decorate, decoration={brace, mirror, amplitude=2mm}]
        (0.02\linewidth,-0.25\linewidth) --  (\linewidth,-0.25\linewidth);

  \node[anchor=south] at (0.135\linewidth,-0.35\linewidth){No clipping};
  \imginfig{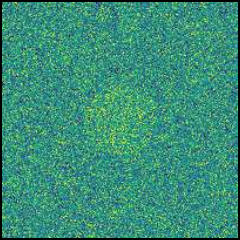}{0.00}{-0.60}{0.24}; \rotatedtxt{Median}{0.01}{-0.47}{white}
  \imginfig{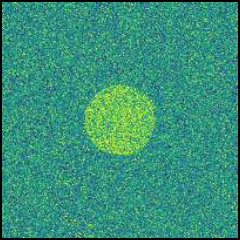}  {0.00}{-0.85}{0.24}; \rotatedtxt{Mean}  {0.01}{-0.71}{white}
  \imginfig{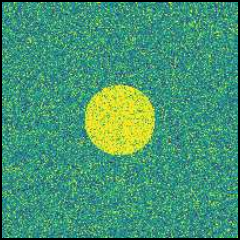}   {0.00}{-1.10}{0.24}; \rotatedtxt{STD}   {0.01}{-0.96}{white}

  \node[anchor=south] at (0.375\linewidth,-0.35\linewidth){$\sigclipsigma$ $\sigma$-clip};
  \imginfig{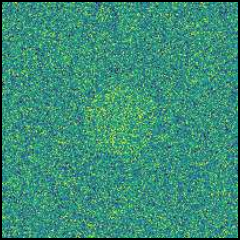}{0.25}{-0.60}{0.24};
  \imginfig{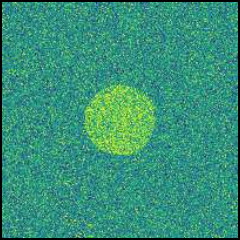}  {0.25}{-0.85}{0.24};
  \imginfig{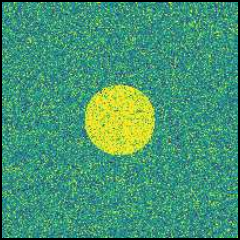}   {0.25}{-1.10}{0.24};
  \imginfig{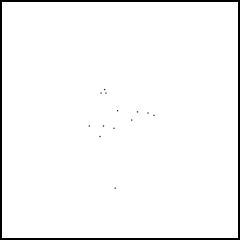}{0.25}{-1.35}{0.24};
  \rotatedtxt{Coadd pixels}{0.10}{-1.21}{black}
  \rotatedtxt{using $<8$}{0.15}{-1.21}{black}
  \rotatedtxt{inputs:}{0.20}{-1.21}{black}

  \node[anchor=south] at (0.63\linewidth,-0.35\linewidth){$\madclipmad$ MAD-clip};
  \imginfig{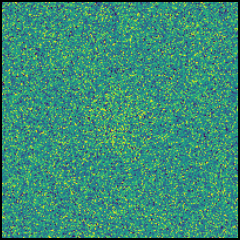}{0.50}{-0.60}{0.24};
  \imginfig{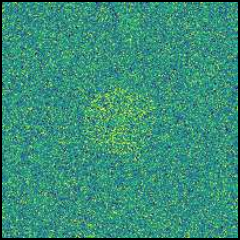}  {0.50}{-0.85}{0.24};
  \imginfig{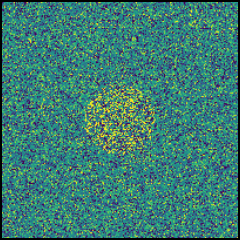}   {0.50}{-1.10}{0.24};
  \imginfig{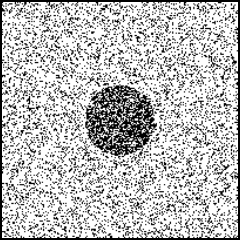}{0.50}{-1.35}{0.24};

  \node[anchor=south] at (0.875\linewidth,-0.35\linewidth){Proposed};
  \imginfig{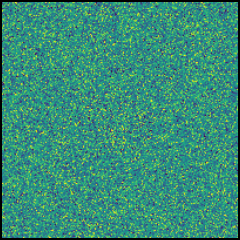}{0.75}{-0.60}{0.24};
  \imginfig{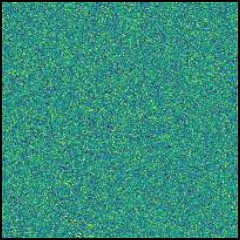}  {0.75}{-0.85}{0.24};
  \imginfig{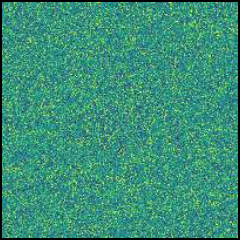}   {0.75}{-1.10}{0.24};
  \imginfig{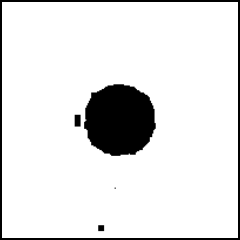}{0.75}{-1.35}{0.24};

  \node [at={(0.006\linewidth,-1.4\linewidth)},
         rectangle,
         text centered,
         font=\ttfamily,
         line width=1pt,
         anchor=north west,
         draw=black!30!white,
         fill=black!05!white,
         rounded corners=0.1cm,
         text width=0.976\linewidth,
         minimum width=0.205\linewidth,
         minimum height=0.26\linewidth,
         label={[shift={(0cm,-6mm)}]Masking outliers in the exposure with an outlier}
         ] {};

  \imginfig{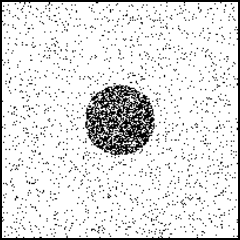}  {0.00}{-1.67}{0.19};
  \imginfig{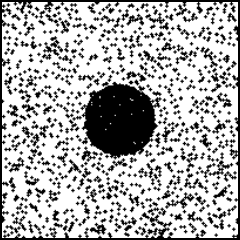}{0.20}{-1.67}{0.19};
  \imginfig{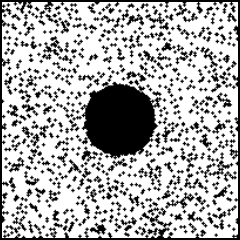}  {0.40}{-1.67}{0.19};
  \imginfig{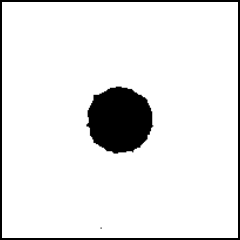}   {0.60}{-1.67}{0.19};
  \imginfig{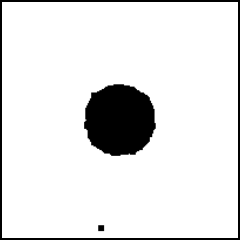}{0.80}{-1.67}{0.19};

\end{tikzpicture}

%% file: tex/build/macros/dependencies.tex
 
This research was done with the following free software programs and libraries: Bzip2 1.0.8, CFITSIO 4.6.3, cURL 8.17.0, Dash 0.5.12, Discoteq flock 0.4.0, Expat 2.6.4, File 5.46, Fontconfig 2.16.0, FreeType 2.13.3, Git 2.52.0, GNU Astronomy Utilities 0.24 \citep{gnuastro}, GNU Autoconf 2.72, GNU Automake 1.17, GNU AWK 5.3.2, GNU Bash 5.3.9, GNU Binutils 2.45.1, GNU Bison 3.8.2, GNU Compiler Collection (GCC) 16.1.0, GNU Coreutils 9.11, GNU Diffutils 3.12, GNU Findutils 4.10.0, GNU gettext 1.0, GNU gperf 3.1, GNU Grep 3.12, GNU Gzip 1.14, GNU Integer Set Library 0.27, GNU libiconv 1.18, GNU Libtool 2.5.4, GNU libunistring 1.4.2, GNU M4 1.4.21, GNU Make 4.4.1, GNU Multiple Precision Arithmetic Library 6.3.0, GNU Multiple Precision Complex library, GNU Multiple Precision Floating-Point Reliably 4.2.2, GNU Nano 9.0, GNU NCURSES 6.5, GNU Readline 8.3.3, GNU Scientific Library 2.8, GNU Sed 4.9, GNU Tar 1.35, GNU Texinfo 7.2, GNU Wget 1.25.0, GNU Which 2.23, GPL Ghostscript 10.06.0, Help2man , IANA Time Zone Database (tzdb) 2026a, Less 685, Libffi 3.4.7, libICE 1.1.2, Libidn 1.42, Libjpeg 9f, Libpaper 1.1.29, Libpng 1.6.46, libpthread-stubs (Xorg) 0.5, libSM 1.2.5, Libtiff 4.7.0, libXau (Xorg) 1.0.12, libxcb (Xorg) 1.17.0, libXdmcp (Xorg) 1.1.5, libXext 1.3.6, Libxml2 2.15.1, libXt 1.3.1, Lzip 1.25, OpenSSL 3.6.0, PatchELF 0.13, Perl 5.42.0, pkg-config 0.29.2, Python 3.13.2, Sqlite 3.53.0, Tcl 9.1a1, util-Linux 2.41.3, util-macros (Xorg) 1.20.2, WCSLIB 8.5, X11 library 1.8, XCB-proto (Xorg) 1.17.0, xorgproto 2024.1, xtrans (Xorg) 1.5.2, XZ Utils 5.6.3 and Zlib 1.3.1. 
The \LaTeX{} source of the paper was compiled to make the PDF using the following packages: courier 77161 (revision), epsf 2.7.4, etoolbox 2.5m, fontaxes 2.0.2, helvetic 77161 (revision), lineno 5.9, newtx 1.756, pgf 3.1.12, pgfplots 1.18.2, revtex4-1 4.1s, tex 3.141592653, textcase 1.05, times 77161 (revision), txfonts 77682 (revision), ulem 78931 (revision), xkeyval 2.10 and xpatch 0.3. 
We are very grateful to all their creators for freely  providing this necessary infrastructure. This research  (and many other projects) would not be possible without  them.